\documentclass{DISproc}

\begin{document}
\title{Overview of Deeply Virtual Compton Scattering at {\sc hermes}}

\author{{\slshape Morgan Murray$^1$}for the H{\sc ermes} collaboration.\\[1ex]
$^1$\,Rm.514b, University of Glasgow, Glasgow, G12 8QQ, Scotland}

\contribID{04}

\doi  

\maketitle

\begin{abstract}
Deeply Virtual Compton Scattering represents the best experimental channel
through which to understand Generalised Parton Distributions. The {\sc hermes}
experiment measured the most diverse set of DVCS results of any experiment;
this talk discusses the most recent sets of DVCS results released by {\sc hermes}
and the unique experimental conditions found at {\sc hermes} that facilitated the
measurements. We also examine the various ways in which the {\sc hermes}
experimental measurements are being used to constrain GPDs and how future
experiments can learn from the {\sc hermes} program.
\end{abstract}

\section{Introduction}

The {\sc hermes} experiment at {\sc desy} ran from 1995 to 2007 as a forward spectrometer on the {\sc hera} electron/positron beam. Its original \emph{raison d'\^etre} was the investigation of the spin structure of the nucleon, a purpose continued in the later years of the experiment's lifetime with its focus on exclusive physics. These proceedings report briefly on selected {\sc hermes} results on Deeply Virtual Compton Scattering which is the leptoproduction of real photons from a nuclear target --- in the case of {\sc hermes} this reduces to electroproduction on typically a Hydrogen target.

The study of exclusive physics is performed with a view to obtaining information on Generalised Parton Distributions (GPDs). These theoretical objects contain a wealth of information on nucleonic structure, including a way to access the total angular momentum of the constituent partons~\cite{Ji} and correlated information pertaining to the distribution of partons in the transverse spatial plane of the nucleon with the fraction of the nucleon's longitudinal momentum carried by that parton~\cite{Burkardt}. There are four distributions of interest that are expected to enter into typical scattering experiments at leading twist with the least kinematical suppression: $H, E, \widetilde{H}$ and $\widetilde{E}$. Generalised Parton Distributions only describe the soft part of the diagram and thus appear in data from lepton scattering experiments convoluted with a hard scattering kernel~\cite{BMK}. The resultant distribution is referred to a Compton Form Factor (CFF) and is typically denoted $\mathcal{H}, \mathcal{E},\mathcal{\widetilde{H}}$ and $\mathcal{\widetilde{E}}$ for GPDs $H, E, \widetilde{H}$ and $\widetilde{E}$ respectively.

The process $ep\rightarrow ep\gamma$ has two contributors; alongside the DVCS process there is a contribution from the Bethe-Heitler process ($ep$ scattering with a Bremsstrahlung photon). The scattering amplitudes from each process interfere and provide the cross-section with three contributory terms: one term from the squared amplitude from each process and one from the interference between the two processes. Typically at the kinematic space covered by {\sc hermes} the BH process dominates but access to the large interference term can provide information derived form CFFs and thus GPDs. Mostly asymmetries are constructed so as to minimise the contribution from the pure BH process.

\section{The HERMES Experiment}
The H{\sc ermes} experiment has been covered in detail in the literature, eg.~\cite{ackerstaff}, and no such detailed description will be repeated here. This work will refer solely to BH/DVCS detected by scattering a positron or electron off a proton target, with the produced photon being detected in the electromagnetic calorimeter. Bethe Heitler/Deeply Virtual Compton Scattering events are traditionally selected at {\sc hermes} by use of a missing mass technique --- the recoiling proton is scattered outside of the H{\sc ermes} acceptance, so the final data sample comprises the desired BH/DVCS candidate events and some background processes~\cite{HERMES1}--\cite{HERMES9}.
 The largest contributor to these background processes is BH involving a proton resonance which typically makes up approx. 10\% of the data sample. The {\sc hermes} spectrometer was upgraded in 2006 with a recoil proton detector in the target region~\cite{recoilTDR} in order to eliminate this background process. The {\sc hermes} set of DVCS results now includes a beam helicity measurement made using this procedure.

\section{Deeply Virtual Compton Scattering}

H{\sc ermes} was designed to measure asymmetries. Asymmetries in the azimuthal distribution ($\phi$) of the produced real photon in DVCS are measured to provide information that can be used to constrain Generalised Parton Distributions~\cite{BMK}. As a consequence of the unique {\sc hermes} setup, there are several different combinations of beam and target states that supplied useful information for DVCS:
\begin{eqnarray}
\mathcal{A}_{\rm{C}}(\phi) &\equiv& \frac{\rm{d}\sigma^+(\phi) - \rm{d}\sigma^-(\phi)} {\rm{d}\sigma^+(\phi) + \rm{d}\sigma^-(\phi)} \label{e:bca}\\
\mathcal{A}_{\rm{LU}}(\phi) &\equiv& \frac{\rm{d}\sigma^\rightarrow (\phi) - \rm{d}\sigma^\leftarrow (\phi)}{\rm{d}\sigma^\rightarrow (\phi) + \rm{d}\sigma^\leftarrow (\phi)}\label{e:bsa}\\
\mathcal{A}_{\rm{UT}}(\phi) &\equiv& \frac{\rm{d}\sigma^\Uparrow (\phi) - \rm{d}\sigma^\Downarrow (\phi)}{\rm{d}\sigma^\Uparrow(\phi) + \rm{d}\sigma^\Downarrow (\phi)}\label{e:ttsa}\\
\mathcal{A}_{\rm{LT}}(\phi) &\equiv& \frac{\big(\rm{d}\sigma^{\stackrel{\rightarrow}{\Uparrow}}(\phi) + \rm{d}\sigma^{\stackrel{\leftarrow}{\Downarrow}}(\phi)\big) -
\big(\rm{d}\sigma^{\stackrel{\rightarrow}{\Downarrow}}(\phi) + \rm{d}\sigma^{\stackrel{\leftarrow}{\Uparrow}}(\phi) \big)}
{\big(\rm{d}\sigma^{\stackrel{\rightarrow}{\Uparrow}}(\phi) + \rm{d}\sigma^{\stackrel{\leftarrow}{\Downarrow}}(\phi)\big) +
\big(\rm{d}\sigma^{\stackrel{\rightarrow}{\Downarrow}}(\phi) + \rm{d}\sigma^{\stackrel{\leftarrow}{\Uparrow}}(\phi)\big)}\label{e:ltsa}\\
\mathcal{A}_{\rm{UL}}(\phi) &\equiv& \frac{\rm{d}\sigma^\Rightarrow (\phi) - \rm{d}\sigma^\Leftarrow (\phi)}{\rm{d}\sigma^\Rightarrow (\phi) + \rm{d}\sigma^\Leftarrow (\phi)}\label{e:aul}\\
\mathcal{A}_{\rm{LL}}(\phi) &\equiv& \frac{\big( \rm{d}\sigma^{\stackrel{\rightarrow}{\Rightarrow}}(\phi) + \rm{d}\sigma^{\stackrel{\leftarrow}{\Leftarrow}}(\phi) \big) - 
\big( \rm{d}\sigma^{\stackrel{\rightarrow}{\Leftarrow}}(\phi) + \rm{d}\sigma^{\stackrel{\leftarrow}{\Rightarrow}}(\phi)\big)}
{\big( \rm{d}\sigma^{\stackrel{\rightarrow}{\Rightarrow}}(\phi) + \rm{d}\sigma^{\stackrel{\leftarrow}{\Leftarrow}}(\phi) \big) + 
\big( \rm{d}\sigma^{\stackrel{\rightarrow}{\Leftarrow}}(\phi) + \rm{d}\sigma^{\stackrel{\leftarrow}{\Rightarrow}}(\phi)\big)}\label{e:all}
\end{eqnarray}
Here $\rightarrow$ ($\leftarrow$) refers to the direction of the beam helicity and $\Uparrow$ ($\Downarrow$, $\Rightarrow$, $\Leftarrow$) refers to the direction of the target helicity. The beam charge asymmetry in Eq.~\ref{e:bca} is expected to be sensitive mostly to the real part of CFF $\mathcal{H}$, the beam helicity asymmetry in Eq.~\ref{e:bsa} is expected to be sensitive mostly to the imaginary part of CFF $\mathcal{H}$, the longitudinal target spin asymmetry in Eq.~\ref{e:aul} provides access to the imaginary part of $\mathcal{\widetilde{H}}$ and the double spin asymmetry in Eq.~\ref{e:all} provides access to the real part of $\mathcal{\widetilde{H}}$. Since H{\sc era} supplies both beam charges, the interference and DVCS contributions to Eq.~\ref{e:bsa} can be extracted separately.

There are currently two approaches for obtaining GPD information from experimental data. One approach is to fit the CFFs from experimental measurements of asymmetries simultaneously, thus revealing information on the CFFs but not the underlying GPDs~\cite{Moutarde2009}. This approach has the advantage that it is fast and easily understandable, requiring no detailed theoretical work on the underlying GPD, but provides detail only on the CFF which cannot be used to extract underlying physical quantities. The second more thorough approach is to postulate GPDs from first principles and work through the detailed calculations to produce predictions for asymmetries. The underlying calculations can then be revised to provide predictions that are more consistent with the observed data~\cite{Kumericki},~\cite{GGL}. There have also recently been explorations into the extraction of GPDs from {\sc hermes} data using neural networks \cite{KMNN}.

\section{DVCS Measurements @ H{\bf {\sc ermes}}}

\begin{figure}
\centering
\includegraphics[width=\textwidth]{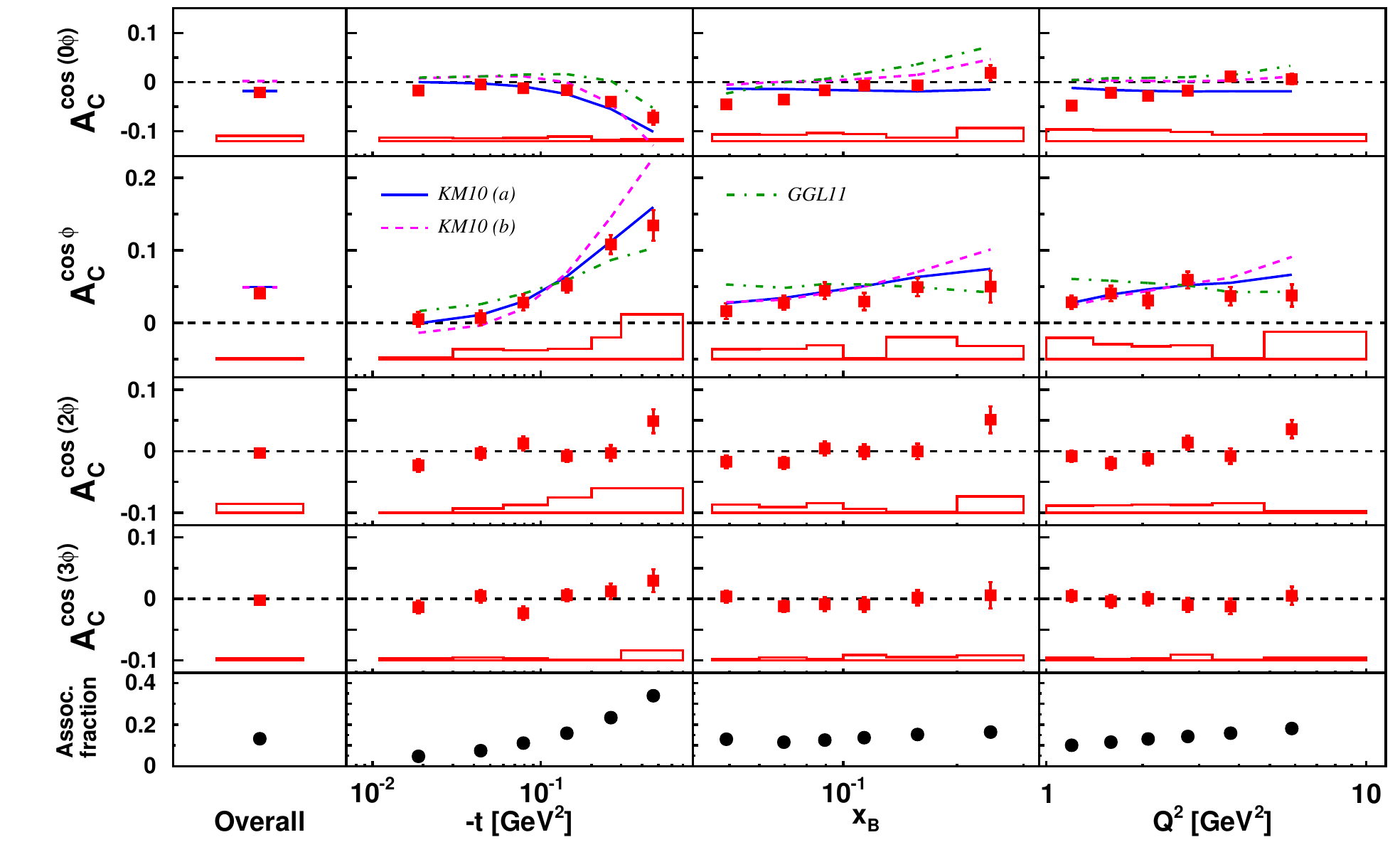}
\includegraphics[width=\textwidth]{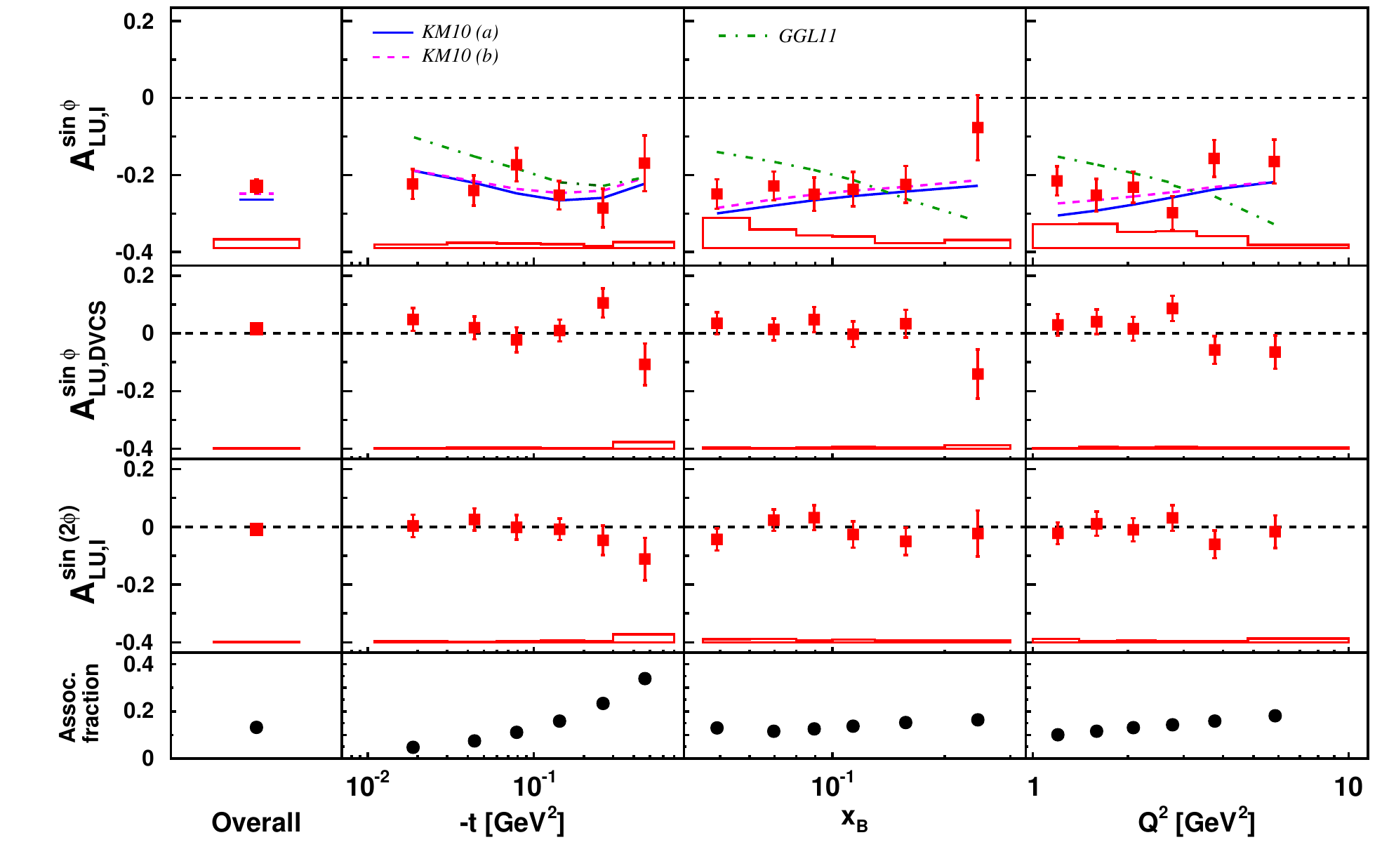}
\caption{The Beam Charge Asymmetry (top) and Beam Helicity Asymmetry (bottom) measured using a missing-mass technique at {\sc hermes}. See text for details.}
\label{f:bas}
\end{figure}

Two of the DVCS measurements made at {\sc hermes} are presented in Fig.~\ref{f:bas}. They correspond to the asymmetries in Eqs.~\ref{e:bca} and~\ref{e:bsa}. The lower panel in each of the figures shows the expected contamination in the data sample from resonance events. Asymmetries with a leading twist contribution from GPD $H$ are given by the $\cos\phi$ harmonic in the upper plot in Fig.~\ref{f:bas} and the $\sin\phi$ harmonic in the lower plot in Fig.~\ref{f:bas}. The other harmonics in the plots show sub-leading twist contributions or harmonics that are expected to be significantly suppressed at {\sc hermes} kinematics. All of these higher-twist amplitudes are compatible with zero.

The beam helicity asymmetry has also been measured at {\sc hermes} using a kinematically complete event selection technique~\cite{recoil}. Extractions of the asymmetry amplitudes of Eqs.~\ref{e:ttsa}, \ref{e:ltsa}, \ref{e:aul} and~\ref{e:all} have also been published by {\sc hermes}~\cite{HERMES3,HERMES7,HERMES9}. The results from the polarised target experiments indicate that there may be non-zero contributions from some transversity and higher-twist GPDs.

{\raggedright
\begin{footnotesize}



\end{footnotesize}
}


\end{document}